\documentclass[showpacs,twocolumn,preprintnumbers,amsmath,amssymb]{revtex4}
\usepackage{graphicx}
\begin{document}
\preprint{IFF-RCA-2011-01}
\title{Is there a Bigger Fix in the Multiverse?}
\author{Pedro F. Gonz\'{a}lez-D\'{\i}az}
\affiliation{Colina de los Chopos, Instituto de F\'{\i}sica
Fundamental,\\ Consejo Superior de Investigaciones
Cient\'{\i}ficas, Serrano 121, 28006 Madrid (SPAIN).}
\date{\today}
\begin{abstract}
The dynamics of baby universes which are branched off in pairs
from a large universe and whose respective members are entangled
to each other has been studied in quantum gravity and string
theory. It is shown that the probability measure for such pairs
essentially keeps its Planckian form when one of their two members
is trapped by a Lorentzian tunnel and travels through it into
another large universe, so that the two baby universes of the
pairs preserve their mutual entanglement even when they are going
to be finally branched in distinct large universes. The conclusion
is thus drawn that big fixing the fundamental constants in our
universe actually big fixes such constants in the set of all
single universes of the multiverse.
\end{abstract}


\maketitle

\vspace{1cm}

One of the reasons why the old ideas of quantum cosmology and
Euclidean quantum gravity have been retaken with renewed interest
in recent years is the growing recognition that the landscape of
the stringy vacua can be too big to allow that reasonable
selection criteria for particle phenomenology and initial
conditions for the universe can be successfully worked out within
string theory itself [1]. Among such ideas there was what Coleman
dubbed the Big Fix [2], according to which the minima of the
Euclidean action provided the most probable values for the
fundamental constant, including a vanishing value for the
cosmological constant. This result corresponded to a wave function
for the universe which was expressed in terms of the exponential
of minus the Euclidean action. That was rather confusing and
counter-intuitive [3] because it predicted higher probability for
the largest universes and, more importantly, because it is now
seen to contradict possible interpretations of the current
accelerating expansion of the universe which can only be
compatible with small but still nonzero value for the cosmological
constant or the energy density of dark energy [4]. At the end of
the day, quantum cosmology expressed in terms of the
Wheeler-DeWitt equation also allows for the possibility that the
wave function of the universe can be described as the exponential
of plus the Euclidean action [5], so opening the room for
conclusions other than attributing a small probability for a large
value of the cosmological constant.

There is still another reason that can contribute to switch on an
additional interest in the Coleman Euclidean program when it is
conveniently supplemented with some extra ingredients. It is that
baby universes - that is the Lorentzian version of Euclidean
wormholes [6] which were explicitly used by Coleman [2] in order
to establish the big fix mechanism leading to the fixing of the
fundamental physical constant - need not be created individually
as it was initially assumed in Coleman dynamics, but they can also
be branched off from our universe in correlated baby-universe
pairs [7] which should only be described by means of a density
matrix [8] rather than a wave function. In this case, there will
always be a non vanishing contribution from a mixed state density
matrix to describe the quantum state of the Euclidean wormholes.
It was shown more than fifteen years ago that in such a case, the
probability measure derived by using quantum gravity [9] and
string theory [10] arguments appears to be free from all confusing
situations and counter-intuitiveness shown by the Coleman case and
actually predicts the result required by the current accelerating
evolution of the universe that the most probable value of the
cosmological constant might be small but always nonzero, and hence
the associated big fix appears to be quite more reliable.

I will first very briefly review here the results derived from
quantum gravity and string theory respectively at Refs. [9] and
[10], in a rather qualitative fashion. In the dilute Euclidean
wormhole approximation [6], the effects produced by a single
wormhole on ordinary matter fields in the asymptotic regions can
be expressed in the path integral for the expectation value of any
given observable $O$ by inserting a factor which, if the Euclidean
wormhole quantum state is given by a wave function, does not
depend on any energy-spectrum characteristic. According to the
formalism put forward by Klebanov, Susskind and Banks [11] we thus
find that, after summing over the number of Euclidean wormholes
(so exponentiating the action integral) and making then the
bi-local action local (in this way entering the Coleman
$\alpha$-parameters), one gets the Coleman simple exponential
probability law [12] that leads to the uncertainties and
shortcomings pointed out before. No further summation should be
here performed as the resulting expression does not depend on any
Euclidean wormhole spectrum or other discrete indices.

However, if the baby universes are branched off and in in pairs,
after performing the above two steps which had to be also
performed in the individual baby universe case (that is, summing
over the number of wormholes and making local the resulting
exponent), then one has [9,10] to also sum over the energy
spectral index $k=m-n\geq 1$ on which the expression resulting
from the two first steps depends. The dynamics is in that case
described in terms of a Planckian-like law [9,10] which gives a
maximal probability of order unity not when the cosmological
constant vanishes but whenever it gets a nonzero value which can
be done so small as the Coleman's alpha parameter allows it. The
fixing of the other constant of physics will similarly be
determined by other minima of the Euclidean action governed by
that Planckian-like probability law. It is worth noticing moreover
that, if we normalize the probability to a maximal value unity,
then smaller universes would be more probable than larger ones,
and hence inflation is predicted to be favored. The quantum
content represented by such a law corresponds to a quantization of
gravity which is over and above that is contained in the usual
formulation of quantum cosmology. It should necessarily be
expressing the only distinctive feature introduced in the
derivation of the law that the two baby universes of the pairs are
correlated to each other. Nucleating a pair of baby universes the
two at once looks quite the same as the emission of two entangled
photons by a nonlinear optical process in EPR experiments. The
main hypothesis of the present paper is to interpret the Planckian
shape of the quantum-gravity probability law as a direct
consequence from an the entanglement between the two baby
universes in each branched off pair, in such a way that if one
determines the state of one of the two baby universes of one such
pairs the state of the other baby universe of the same pair is
automatically determined as well, even in the case that we do not
perform any measurement on the latter baby universe.

In what follows we shall consider a framework in which an
entangled baby universe pair is branched off from a parent
universe, denoted Universe I (see Fig. 1), of which just one of
the baby universes is trapped by a Lorentzian space-time tunnel
through which it travels into another parent universe, denoted
Universe II (see Fig. 1), at which it is finally branched in,
leaving the tunnel, while the non trapped baby universe of the
original pair branches in back in parent universe I. Any baby
universe that enters a tunnel should be expected to couple with
the tunnel as a whole from one mouth to the other, and the baby
universe-tunnel coupling would only depend on the kind of
tunneling - whether it is a wormhole, a ringhole or a Klein-bottle
hole - so as on the relative velocity of its two mouths and the
physical parameters and topology that define its throat. When one
allows the two wormhole mouths to move relative to each other with
a nonzero velocity, then the trapped baby universe would be
expected to travel in time.

One expects that any of these coupling and mouth vibrating
processes will generally take place without any relevant change in
the mutual entanglement between the two baby universes. The
physical reason for that is that the two compact, closed
constructs branched off from the parent universe do not originally
evolve in any background Lorentzian space-time, So that their
mutual entanglement cannot be defined to depend on any background
Lorentzian space-time. Thus, the space-time placement of the baby
universe along the Lorentzian wormhole tunneling cannot be
involved at the two baby universe mutual entanglement. Therefore,
the unique blueprint that any traveling of one baby universe along
the tunnel would leave in the probability measure defined in terms
of a path integral would just be in the form of a given overall
coupling constant $\xi_j$ (if it is baby universe $j$ which
couples to the Lorentzian tunnel, see Fig. 1) entering the action
integral, and this modification implies no further summation or
integration (tracing off) requirement for deriving the dynamics,
and hence the resulting probability expression should keep its
Planckian shape.

In order to work out the precise form of the dynamics and derive
from it the physical implications for the situation depicted in
Fig. 1, we shall proceed as follows. In the dilute Euclidean
wormhole approximation [6,12], the effects produced by Euclidean
wormholes on ordinary matter fields on the asymptotic regions of
Universe I can be expressed in the path integral for the
expectation value of a given observable $O$ by a double insertion:
a factor which, if the Euclidean wormhole quantum state is given
by a density matrix, is
$-\frac{1}{2}C_{ij}\beta_i\beta_j\epsilon_{mn}^{-1}$, and a factor
$\gamma_j=\xi_j^{-1}$ inversely depending on the overall coupling
of the Lorentzian baby universe $j$ to the Lorentzian wormhole
(see Fig. 1). We then have [9]
\begin{equation}
\langle O\rangle=\int dgOe^{-I(g,\lambda)}\left(\frac{1}{2}
\sum_{i,j}C_{ij}\gamma_{j}(y)\beta_i(x)\beta_j(y)\epsilon_{mn}^{-1}\right)
,
\end{equation}
where $C_{ij}\propto e^{-S_w}=D_{ij}^{-1}$, with $S_w$ the
Euclidean action for the Euclidean wormhole,
$\epsilon_{mn}=E_{m}^{(f)}-E_n^{(g)}$, the $E$'s being the energy
levels for the matter fields ($f$) and the gravitational field
($g$) harmonic oscillators, respectively, and $\epsilon_{mn}^{-1}$
the relative probability for the given state $\Psi_{mn}$ [8].
$\lambda$ collectively denotes parameters such as coupling
constants, particle masses, the cosmological constant, etc.;
$\beta_i=\frac{1}{\sqrt{2}}\int d^4 x\sqrt{g(x)}V_i(x)$, with
$V_i$ denoting the vertex operator and index $i$ labeling the
elements of the basis for the local field operators at the point
$x$ on the large region. Indices $i,j$ are independent of the
quantum numbers $n,m$ which label the off shell-energy spectrum.
All dependence of the path integral on that spectrum and on the
coupling of the baby universe to a Lorentzian tunnel are
incorporated through the relative probability factor
$\epsilon_{mn}^{-1}$ and the factor $\gamma_j$, respectively,
because the respective unique invariant theories for
asymptotically flat Euclidean wormholes and asymptotically flat
space-time tunnels must satisfy the boundary requirement that all
possible kinds of Euclidean wormholes and Euclidean wormhole
states, on the one hand, and Lorentzian tunnels and its possible
classical or even quantum states, on the other hand, ought all to
have equal asymptotic behavior, respectively, thus rendering the
insertion amplitude to join their ends onto the asymptotic regions
independent of the Euclidean wormhole spectrum and the baby
universe-Lorentzian wormhole coupling. The dependence on the
eigenenergies $\epsilon_{mn}$ arises because the Euclidean
wormhole is off-shell in the case that the baby universes are
nucleated in pairs. If such universes are Planck-sized, then we
have the most probable case where $\epsilon_{m,n}=m-n$. Now,
following Coleman [2], we first sum over any number of Euclidean
wormholes, and then make local the resulting exponent, to get
\begin{equation}
\int dg O e^{-I(g,\lambda)}\int\prod_p d\alpha_p
e^{-\frac{1}{2}(m-n)D_{ij}\xi_j\alpha_i\alpha_j}e^{-\beta_l\alpha_i}
,
\end{equation}
through which the position-independent parameters $\alpha$'s enter
the formalism.

Now, unlike the case considered by Coleman [2,12], we should next
sum [9,10] over all possible Euclidean wormhole states. Avoiding
the over counting that comes from summing over $n$ and $m$
independently, so as restricting to indices such that $m>n$, in
order to ensure a converging path integral, and taking the lower
limit in the sum to be unity, in order to keep the off-shell
character even classically, we finally obtain
\begin{equation}
\langle O\rangle\propto \int d\alpha P(\alpha)Z(\alpha)\langle
O\rangle_{\lambda+\alpha} ,
\end{equation}
where the third exponent of the previous expression has been
inserted in the path integral of the parent universe I, and
\begin{equation}
P(\alpha)=\frac{1}{e^{D_{ij}\xi_j \alpha_i\alpha_j}-1},\;\;\;
Z(\alpha)=\int dge^{-I(g,\lambda+\alpha)} .
\end{equation}

\begin{figure}
\includegraphics[width=.9\columnwidth]{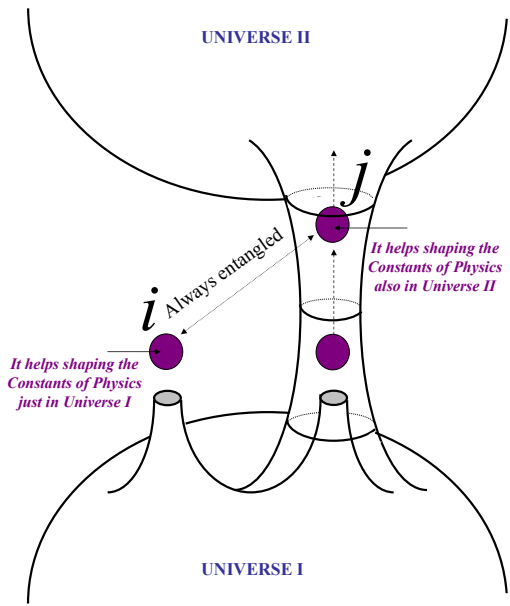}
\caption{\label{fig:epsart} Pictorial representation of the
nucleation of an entangled baby universe pair in the context of
the multiverse, with one of the baby universes branching off
freely from the parent universe, and the other being trapped by a
Lorentzian tunnel (a wormhole, a ringhole or a klein-bottle hole)
along which it may travel to the other universe. After a short
time the baby universe trapped in the Lorentzian tunnel arrives at
and is branched in the universe other than the one where it was
originally branched off, time traveling or not. The important
result is that the entanglement between the two baby universes is
expected to be preserved in both cases, so that determining the
values of the physical constants of Universe I according to the
improved Coleman's big fix mechanism considered here immediately
determines the values of those constants of Universe II.}
\end{figure}

These results can be rather straightforwardly confirmed in the
two-dimensional realm of string theory. In fact, one can study
Euclidean multi-wormhole configurations in Polyakov stringy theory
by looking at the Euclidean wormholes as [10,13] the handles on a
Riemann surface of genus $\rho$, with $\rho$ the number of handles
or Euclidean wormholes in the configuration in the case depicted
in Fig. 1. Starting thus with the Green function that describes
the effects of the handles on first order tachyonic amplitudes
taken as the path integral over all space-time coordinates on the
Riemann surface, and by following the calculation performed in
Ref. [10], supplemented by the arguments on the space-time
independence of the coupling constant between the baby universes
and the Lorentzian tunnels, stated before, we finally obtain the
same result as from quantum gravity, provided we express the Green
function in terms of the handle quantum state on the circle using
the Fourier transform of the delta function [13] (ensuring that,
after cutting the handles, points on a resulting circle are
identified with points on the other resulting circle) for the zero
mode, and then expanding the delta function for the nonzero modes
in terms of the complete set of the orthonormal
harmonic-oscillator eigenstates that are the solutions of the
string analogous of the Wheeler-DeWitt equation [13]. It can be
shown that this procedure leads to a bi-local effective action
given by
\begin{equation}
-\int d\sigma_1 \int d\sigma_2\sum\int d^4 K\kappa\xi_2
V_p(\sigma_1) V_p(\sigma_2),
\end{equation}
where the $V_p$'s are the handle vertex operators for the fields
on the two circles, $\kappa=\left(K^2
+\sum|n|m_n^{(i)}-2\right)^{-1}$, which results from integrating
over the fields, with $K$ the momentum for the zero mode $n=0$,
and $m_n$ labeling the excited states of the resulting
Wheeler-DeWitt wave function [13], $e^{-\frac{1}{2}n(Y_n^{(i)})^2}
H_{m_n^{(i)}} (\sqrt{n}Y_n^{(i)})e^{iKx_0}$, with $i=1,2$,
$Y_n^{(1)}=\frac{1}{2}(x_n +x_{-n})$, $Y_n^{(2)}=\frac{1}{2i}(x_n
-x_{-n})$, and the coupling constant $\xi_2$ means that we have
chosen position 2 to be occupied by the baby universe which enters
the Lorentzian tunnel. We then first convert action (5) into a
local quantity by inserting the Coleman parameters, summing then
over all $m_n$, to finally have [10]
\begin{equation}
Z(\alpha)\prod_q
\left(e^{\frac{1}{2}n^2\xi_2\alpha_q^2}-1\right)^{-1} ,
\end{equation}
which, when one identifies $D_{ij}\equiv n^2$ and $2\equiv j$, and
takes into account that, out from all possible combinations of
indices $i,j$, only the diagonal combination $i=j=q$ is allowed by
quantum requirements [10], becomes fully equivalent to Eqns. (3)
and (4) which was obtained in the context of four-dimensional
quantum gravity.

If the different single universes of the multiverse are mutually
interconnected to each other by means of space-time tunnels, such
as it must be expected to happen, there then will always be a
nonzero contribution to the path integral describing the state of
the multiverse from branched off baby universe pairs so that that
there always be individual baby universes traveling through
Lorentzian tunnels which are entangled to their partner in the
parent universe in which they were both originally branched off as
an entangled pair, so that all universes in the multiverse are
connected to each other in such a way that if you determine by
means of measurements the values the fundamental constants in a
given universe you are at the same time determining the value of
such constants in all other universes in the multiverse. We obtain
thus a bigger fix for the fundamental constants both in our
universe and in the rest of universes of the considered
multiverse.

On the other hand, from inspection of Eqns (3) and (6) we can
deduce:
\begin{itemize}
\item Entanglement between two baby universes is always preserved
unless for the extreme case in which one of them very strongly
couples to a space-time tunnel.
\item If such a coupling is weak enough then the average number of
existing baby universe pairs should tend to be larger so that the
whole set tends toward its classical limit.
\item If the coupling is strong then the average number of
existing baby universe pairs should tend to be small and separates
from the classical behavior.
\item Finally, if the two Lorentzian tunnel mouths move with
velocity $v$ relative to one another, then the above coupling
would be expected to depend on $v$ in such a way that the larger
$v$ the larger $\xi$. One thus might tentatively assume
$\xi\propto\frac{2\xi^{(o)}}{2-v}$. It would then follow that time
traveling of one of the baby universes of a pair can never
completely destroy mutual entanglement between the two baby
universes.
\end{itemize}

The main conclusion coming from the above calculations and the
consequences drawn from them is that when one settles the values
of the fundamental constants in a given single universe, such
constants and laws should also be settled down once and for all in
the rest of single universes of the multiverse. It follows that,
in spite of the classical mutual independence of the single
universes of one another, if the existence of some classical
tunneling mechanism connecting the distinct single universes
exhaustively among them is allowed, and the nucleation of mutually
entangled baby universe pairs is permitted, with one of these baby
universes being trapped by the tunneling network mechanism, then
it is possible to conceive lab experiments or physical
observations on Earth which will check the physical existence of
other single universes in the multiverse.

\acknowledgements The author thanks Carmen L. Sig\"{u}enza for
useful comments and discussion. This work was supported by the
Spanish MICINN under Research Project No. FIS2008-06332 and is
herewith dedicated to Guillermo who was born at the time the paper
was completed.

\end{document}